\documentclass[]{spie}  
\usepackage[]{graphicx}

\title{Towards a reliability assessment of the c$^{2}_{N}$ and wind speed vertical profiles retrieved from GeMS} 

\author{Benoit Neichel\supit{a}, Elena Masciadri\supit{b}, Andr\`es R. Guesalaga\supit{c}, Franck Lascaux\supit{b} and  Cl\'ementine B\'echet\supit{c}
\skiplinehalf
\supit{a}Aix Marseille Université, CNRS, LAM (Laboratoire d'Astrophysique de Marseille) UMR 7326, 13388, Marseille, France; \\
\supit{b} INAF - Osservatorio Astrofisico di Arcetri, Italy;\\
\supit{c} Pontificia Universidad Cat\'olica de Chile, 4860 Vicu\~na Mackenna, Casilla 7820436, Santiago, Chile 
}

\authorinfo{Send correspondence to Benoit Neichel (benoit.neichel@lam.fr)}

\begin{document} 
  \maketitle 

\begin{abstract}
The advent of a new generation of Adaptive Optics systems called Wide Field AO (WFAO) mark the beginning of a new era. By using multiple Guide Stars (GSs), either Laser Guide Stars (LGSs) or Natural Guide Stars (NGSs), WFAO significantly increases the field of view of the AO-corrected images, and the fraction of the sky that can benefit from such correction. Different typologies of WFAO have been studied over the past years. They all require multiple GSs to perform a tomographic analysis of the atmospheric turbulence. One of the fundamental aspects of the new WFAO systems is the knowledge of the spatio-temporal distribution of the turbulence above the telescope. One way to get to this information is to use the telemetry data provided by the WFAO system itself. Indeed, it has been demonstrated that WFAO systems allows one to derive the C$_{N}^{2}$ and wind profile in the main turbulence layers (see e.g. Cortes et al. 2012\cite{cortes2012atmospheric}). This method has the evident advantage to provide information on the turbulence stratification that effectively affects the AO system, property more difficultly respected by independently vertical profilers. 
In this paper, we compare the wind speeds profiles of GeMS with those predicted by a non-hydrostatical mesoscale atmospherical model (Meso-NH). It has been proved (Masciadri et al., 2013\cite{elena2013meso}), indeed, that this model is able to provide reliable wind speed profiles on the whole troposphere and stratosphere (up to 20-25 km) above top-level astronomical sites. Correlation with measurements revealed to be very satisfactory when the model performances are analyzed from a statistical point of view as well on individual nights. Such a system appears therefore as an interesting reference to be used to quantify the GeMS wind speed profiles reliability. 
\end{abstract}



\section{Introduction}
\label{sec:intro}  

\subsection{From Single Conjugate Adaptive Optics to Wide-Field Adaptive Optics}
Adaptive Optics (AO) is a technique that aims at compensating quickly-varying optical aberrations to restore the ultimate angular resolution limit of an optical system. It uses a combination of Wave-Front Sensors (WFSs), to analyse the light wave aberrations, and Deformable Mirrors (DMs) to compensate them. In the early nineties, astronomers experimented with the technique with the goal of overcoming the natural "seeing" frontier, the blurring of images imposed by atmospheric turbulence\cite{roddier1981effects}. This latter restricts the angular resolution of ground-based telescope to that achievable by a 10 to 50 cm telescope, an order of magnitude below the diffraction limit of 8-m class telescopes. Although successful, two main limitations reduced the usefulness of AO and its wide adoption by the astronomical community. First, the need for a bright guide star to measure the wave-front aberrations and second, the small field of view compensated around this guide star, typically a few tens of arcseconds. The first limitation was solved by creating artificial guide stars, using lasers tuned at 589nm, which excite sodium atoms located in the mesosphere around 100km altitude \cite{foy1985feasibility}. These Laser Guide Stars (LGS) could be created at arbitrary locations in the sky, thus solving the problem of scarcity of suitable guide stars. Nowadays, all of the major ground based telescopes are equipped with such lasers\cite{wizinowich2012progress}. The second limitation arises from the fact that the atmospheric turbulence is not concentrated within a single layer but spread in a volume, typically the first 15km above sea level. This vertical distribution of the turbulence creates the so-called anisoplanatism effect, that severely limits the AO-corrected field, and the sky coverage. For a typical good astronomical site, the AO corrected field is about a few tens of arcseconds in the Near-Infra Red (NIR). The advent of a new generation of AO systems called Wide Field AO (WFAO) mark the beginning of a new era. By using multiple GSs, either LGS or Natural Guide Stars (NGSs), WFAO significantly increases the field of view (FoV) of the AO-corrected images, and the fraction of the sky that can benefit from such correction. In WFAO systems, the light from several GSs is used to probe the instantaneous three-dimensional phase perturbations, and the turbulence volume is then reconstructed by solving an inverse problem. This technique, called atmospheric tomography, was first introduced by Tallon \& Foy\cite{tallon1990adaptive} and later improved by Johnston \&  Welsh\cite{johnston1994analysis}, Ellerbroek\cite{ellerbroek1994first}, and Fusco et al.\cite{fusco2011mcao}, among others.

For the next generation of Extremely Large Telescopes (ELTs), several wide-field adaptive optics (WFAO) concepts such as ground-layer AO (GLAO), multi-conjuguate AO (MCAO), laser tomography AO (LTAO), and multi- object AO (MOAO) have been proposed. These systems will play a crucial role, since most of them could be used as first-light instruments.

\subsection{The need for accurate profiles}
All the WFAO systems require an estimate of the three-dimensional turbulence volume. The tomographic reconstruction error depends only on the GSs’ constellation, atmospheric conditions, and WFSs characteristics. The geometry of the system sets the first two fundamental limitations of the tomography: unseen modes and unseen turbulence. In addition to this fundamental limitation, an additional term comes due to model/statistical errors. This term is introduced when a faulty knowledge of system and atmospheric conditions is used, which limits the tomographic reconstruction (see Neichel et al. \cite{neichel2008josa}).

For instance, an error on the altitude where the layers are reconstructed affects the frequencies where unseen modes should be regularized/truncated. In fact, the altitudes of the reconstructed layers introduced in the model set the sensitivity of the reconstructors to unseen frequencies. A wrong geometry implies that wrong modes are going to be regularized/truncated. For the frequencies that correspond to well-seen frequencies for the model, the direct inverse of the interaction matrix is performed, whereas these frequencies could correspond to badly seen modes in the real geometry. For these frequencies, an over-amplification of noise could appear. Conversely, some frequencies that are well seen by the real geometry could be treated as unseen frequencies by the model. For these modes, whereas a direct inverse would have worked well, over-regularization or truncation limits the accuracy of the reconstruction. The impact of the model error has been studied in different papers. 
For instance, Conan et al.\cite{conan2001mcao} and Tokovinin \& Viard\cite{toko2001tomo} have studied the sensitivity of the reconstruction error to an error on the C$_{N}^{2}$ profile; LeLouarn and Tallon\cite{miska2001tomo} have explored the influence of the reconstructed turbulent layers’ altitudes compared with the real turbulence volume; Fusco et al.\cite{fusco99phase} have studied the effect of a reconstruction on a limited number of equivalent layers (ELs); Costille \& Fusco\cite{costille2012cn2} studied the impact of sampling the input turbulence for an ELT configuration. All these studies suggest that this error term affects the tomographic reconstruction for large-FoV configurations and that either a good knowledge of the turbulence profile or the use of more GSs to reduce the unseen-frequencies area is required to limit the effect of this model error. The tomographic reconstruction is very sensitive to a model error on layer altitudes, particularly when the FoV covering the GSs constellation increases. A good knowledge of the turbulence profile is then a fundamental step of any WFAO systems.



As a second step to improve the system performance, one could use the wind intensity and direction in each layer if available. As a first order, wind intensity and direction can be used to implement predictive control. In astronomical AO systems, there is a time delay between the start of the WFS measurement and the response of the DM to the command computed from these measurements. This delay reduces the performance of the AO loop in what is known as the temporal error caused by the evolution of phase during this delay. Predictive control (see for instance Gavel and Wiberg\cite{gavel2002prediction}; Le Roux et al.\cite{leroux2004prediction}; Hinnen et al.\cite{hinnen2010prediction} and Poyneer et al.\cite{poyneer2007prediction}) can improve the performance of AO systems substantially if the frozen flow assumption holds and it actually contributes to the phase aberration at the different layers.
Recently, Ammons et al.\cite{ammons2012tomo} also proposed to use the wind profile information in order to reduce the resulting tomographic error.  In addition to reducing temporal error budget terms, they show that there are potentially additional benefits for tomographic AO systems: the combination of wind velocity information and phase height information from multiple guide stars breaks inherent degeneracies in volumetric tomographic reconstruction, producing a reduction in the geometric tomographic error.

\subsection{GeMS}

GeMS, the Gemini MCAO System, is the first multi-sodium based LGS AO system used for astronomy\cite{rigaut2014gems,neichel2014gems}. It uses five LGSs distributed on a 1 arcmin constellation to measure and compensate for atmospheric distortions and delivers a uniform, close to diffraction-limited Near-Infrared (NIR) image over an extended FoV of 2 arcmin. GeMS is a facility instrument for the Gemini South (Chile) telescope, and as such is open for the community. It has been designed to feed two science instruments: GSAOI\cite{mcgregor2004gemini}, a 4k x 4k NIR imager and Flamingo2\cite{elston2003performance}, a NIR multi-object spectrograph. 

\subsection{C$_{N}^{2}$ and wind profile measurements with GeMS} 
Turbulence profiling can be obtained from Shack-Hartmann data by SLODAR-like methods\cite{wilson2002slodar}. These methods are based on optical triangulation for the measurement of the atmospheric optical turbulence profile, using the spatial or spatio-temporal cross-correlations of the slopes measured by the WFSs, each pointing at different guide stars. The turbulence strength can be estimated in as many altitude bins as subapertures across the WFSs exist and the SLODAR method can be adapted to use Laser Guide Stars (LGS) as previously reported\cite{gilles2010cn2,cortes2012atmospheric,osborn2012neuro}. In such cases, the turbulence profiling with LGS is performed to non-equally spaced bin altitudes due to the cone effect. As an extension to the SLODAR method, the wind profiling consists in performing time-delayed cross-correlations between all possible combinations of wavefront measurements from the available WFSs data\cite{wang2008cn2}. This profiling method provides information not only on the turbulence distribution in altitude but also on the wind velocity and direction for each layer and their dynamic evolution. The method has also been modified to include multiple LGS-WFSs and the handling of the cone and fratricide effects in the GeMS configuration\cite{cortes2012atmospheric,andres2014cn2}. The method has also been extended to the ESO Adaptive Optics Facility (AOF - see Valenzuela et al.\cite{valenzuela2014spie}) and being used to derive statistical information about Cerro Pach\'on turbulence (see Rodriguez et al.\cite{ignacio2014spie}).

\subsection{Framework of the study}
The overall goal of the study is to asse and quantify the accuracy of the GeMS profiler, in term of C$_{N}^{2}$, wind intensity and wind direction profiles. As a first step toward this objective, we aim at cross-comparing the wind speed of remarkable layers measured by GeMS with those predicted by an atmospherical mesoscale numerical models. In this paper we focus our attention on the wind speed intensity only. Recently, Masciadri et al.\cite{elena2013meso} showed that predicted non-hydrostatical mesoscale atmospherical model were able to provide reliable wind speed profiles on the whole troposphere and stratosphere (up to 20-25 km) above top-level astronomical sites. The statistical estimates done comparing model prediction with 50 radiosoundings distributed in different periods of the year indicate in the range 5-15~km a bias within 1~m$\cdot$s$^{-1}$ and a RMSE within 3 m$\cdot$s$^{-1}$. Above 15~km and in the 3-5~km range, the bias is within 2 m$\cdot$s$^{-1}$. A detailed analysis done on all individual 50 cases (Masciadri et al. 2013\cite{elena2013meso} - Fig.B1) tells us that an equivalent good performance is achieved by the model and that this model can be used as a reference to check/validate GeMS estimates. In this paper, we hence focus on a comparison of wind speed delivered respectively by GeMS and the Meso-NH model. This work is a first attempt in the cross-calibration of GeMS and Meso-NH models results. A more comprehensive cross-calibration is under development. Future work is described in Sect. \ref{sec:perspectives}.

\subsection{GeMS Data Analysis}

\begin{figure}
  \includegraphics[width = 1.0\linewidth]{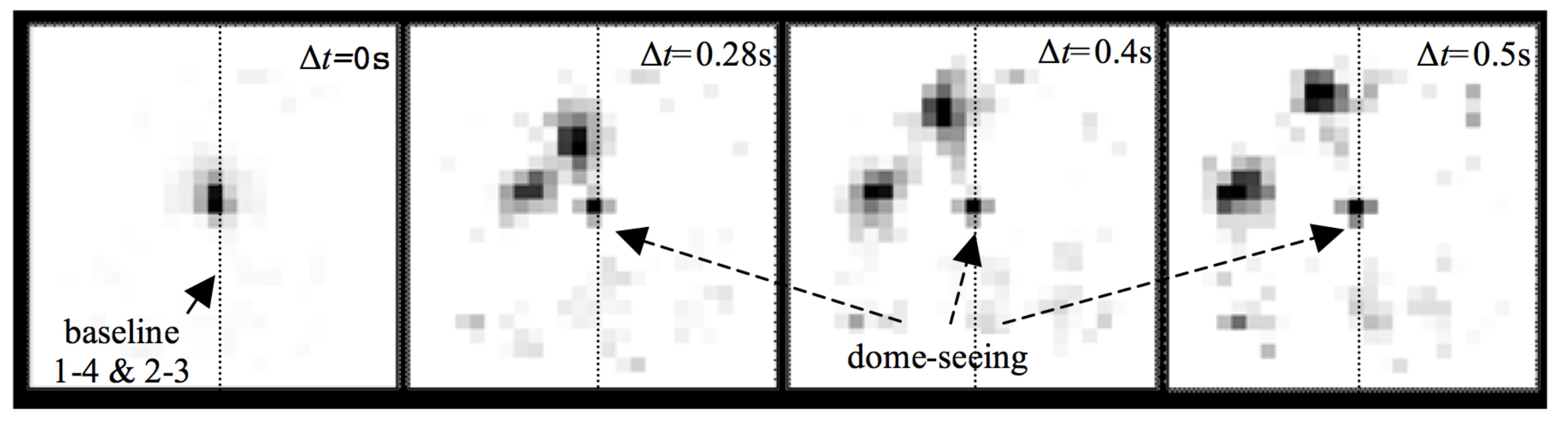}
  \caption[] {The sequence shows the time correlation from $\Delta$t = 0 to $\Delta$t = 0.5s, with two layers moving in different directions and a third static peak at the centre, corresponding to the dome seeing.}
  \label{fig:movie}
\end{figure}

Data analysis is performed following the methods developed in Cortes et al. 2012\cite{cortes2012atmospheric} and Guesalaga et al. 2014\cite{andres2014cn2}. For each telemetry file, we build a C$_{N}^{2}$ profile, and a cross-correlation temporal movie. Fig. \ref{fig:movie} shows a sequence of images extracted from one of this temporal movies. By tracking the correlation peaks we can estimate the direction and speed of the wind, but also the C$_{N}^{2}$ value of the associated layer. We use both the high-resolution and low-resolution profiles to detect the peaks. The peaks are identified by eye, and the wind speed is deduced from the motion of the correlation peak over a given $\Delta$t. Uncertainties in the wind speed intensity are estimated assuming $\pm$0.2 to 0.5 subaperture error, depending on the signal to noise ratio of the peak, and assuming $\pm$ 1 to 2 frames error depending on the translation speed of the peak. Uncertainties in altitude are estimated based on the altitude bin size, as described in Cortes et al. 2012\cite{cortes2012atmospheric}.

\subsection{GeMS Data selection}
The selection criterium is defined as follows. We select GeMS WFS data for which several layers can be clearly and unambiguously detected. Indeed, one of the key difficulty in using WFS data to build wind profiles comes from the identification and the tracking of the layers. Isolating the correlation peaks for tracking purposes is not trivial and it only works in situations where they can be individualized i.e. they have different velocity vectors and are not overlapped along the tracks. It is not the purpose of this paper to present a method for systematic identification of all wind-driven layers. We then selected cases where the cross-correlation peaks are easily traceable. The data-set used in this paper is summarized in Fig. \ref{fig:windprofile1} in the figure titles. The vertical black lines show the time at which GeMS data are available as indicated in the telemetry data in UT.


\section{Wind speed profiles from the atmospherical non-hydrostatic mesoscale model Meso-Nh} 
\label{sec:models}  

All the predictions of atmospheric parameters (wind speed and direction) shown in this study have been obtained with an atmospherical non-hydrostatic mesoscale model called Meso-Nh (Lafore et al.\cite{lafore1998}). Such a model can simulate the temporal evolution of three-dimensional meteorological parameters over a selected finite area of the globe. We refer the reader to Masciadri et al. 2013 \cite{elena2013meso} for the details of the model specification such as temporal scheme, turbulence closure scheme and surface exchanges calculations. The model has been developed by the Centre National des Recherches M\'et\'eorologiques (CNRM) and Laboratoire d'A\'ereologie (LA) de l'Universit\'e Paul Sabatier (Toulouse). 
The package of the optical turbulence ($C_{N}^{2}$) and integrated derived parameters (called Astro-Meso-Nh) have been developed by Masciadri et al. \cite{masciadri1999}. Since then the Astro-Meso-Nh code has been improved along the years by the authors in terms of flexibility providing a set of further outputs and 
permitting us to carry out different scientific studies mainly addressing the reliability of this technique in astronomical applications \cite{masciadri2004, masciadri2006, hagelin2010, hagelin2011, lascaux2012}. 

Both the Meso-Nh code as well as the Astro-Meso-NH code for the optical turbulence are parallelized with OPEN-MPI-1.4.3.
The model can therefore be run on local workstations as well as on the High Performance Computing Facilities (HPCF) cluster of the 
ECMWF, in parallel mode so as to gain in computing time.

\begin{figure}
   \includegraphics[width = 1.0\linewidth]{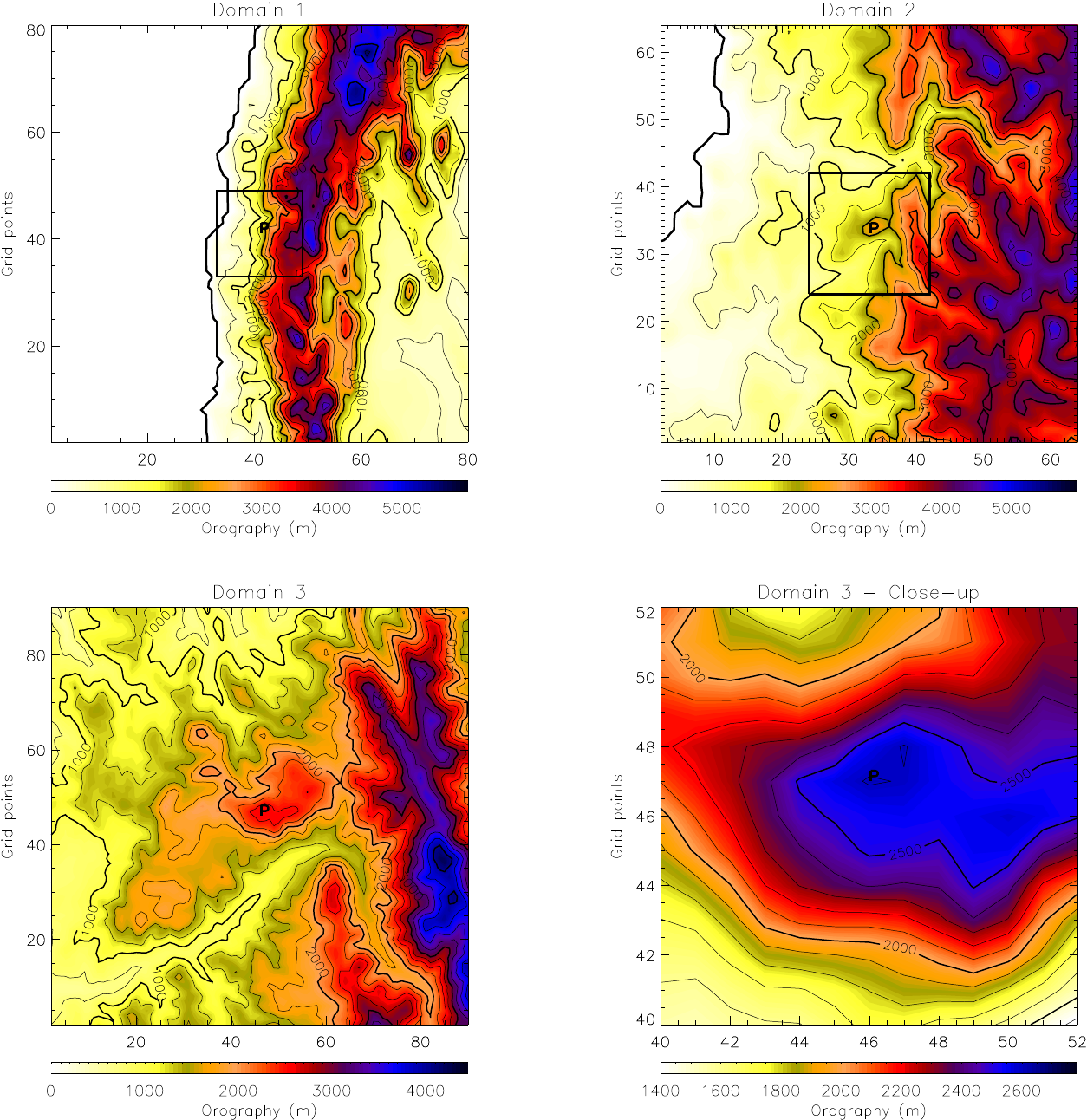}
    \caption[] {Orography. Domain 1: 80x80 grid points; 100 kmx100 km; delta(X)=10 km. Domain 2: 64x64 grid points; 160 kmx160 km; delta(X)=2.5 km. Domain 3: 90x90 grid points; 45 kmx45 km; delta(X)=0.5 km. Letter "P" indicates the location of Cerro Pachon.}
  \label{fig:orography}
\end{figure}

The geographic coordinates of Cerro Pachon are (70$^{\circ}$ 44' 12.096'' W; 30$^{\circ}$ 14' 26.700" S). We used the grid-nesting technique to join the atmospherical model with the topography. Such a techniques consists  of using different imbricated domains of the Digital Elevation Models (DEM i.e orography) extended on smaller and smaller surfaces, with increasing horizontal
resolution but with the same vertical grid. For this study we choose a model configuration with three domains (Fig.~\ref{fig:orography} and Table~\ref{tab:orog}) where the innermost resolution is $\Delta$X~=~500~m. The DEMs we used for this project are the GTOPO\footnote{$http$:$//www1.gsi.go.jp/geowww/globalmap$-$gsi/gtopo30/gtopo30.html$}
with an intrinsic horizontal resolution of 1~km (used for the domains 1 and 2) and SRTM \footnote{http://srtm.csi.cgiar.org/} with an intrinsic horizontal resolution of 90~m for the domain 3. The atmospherical model has been interpolated with the DEM using the horizontal resolutions described in Table \ref{tab:orog}. Fig. \ref{fig:orography} shows the sequence of imbricated DEMs. 

\begin{table}
\label{tab:orog}
\begin{center}
\begin{tabular}{|c|c|c|c|}
\hline
Domain & Grid   & Domain size & $\Delta$X \\
       & Points & (km)        &  (km)  \\
\hline
Domain 1 &  80$\times$80  & 800$\times$800 & $\Delta$X = 10   \\
Domain 2 &  64$\times$64  & 160$\times$160 & $\Delta$X = 2.5 \\
Domain 3 & 90$\times$90 &  45$\times$45  & $\Delta$X = 0.5   \\
\hline
\end{tabular}
\caption{Meso-NH model grid-nesting configuration.
The second column shows the number of horizontal grid-points, the third column the domain extension for each imbricated domain and
the fourth column the horizontal resolution $\Delta$X.}
\end{center}
\end{table}

\begin{figure}
  \begin{tabular}{cc} 
  \includegraphics[width = 0.48\linewidth]{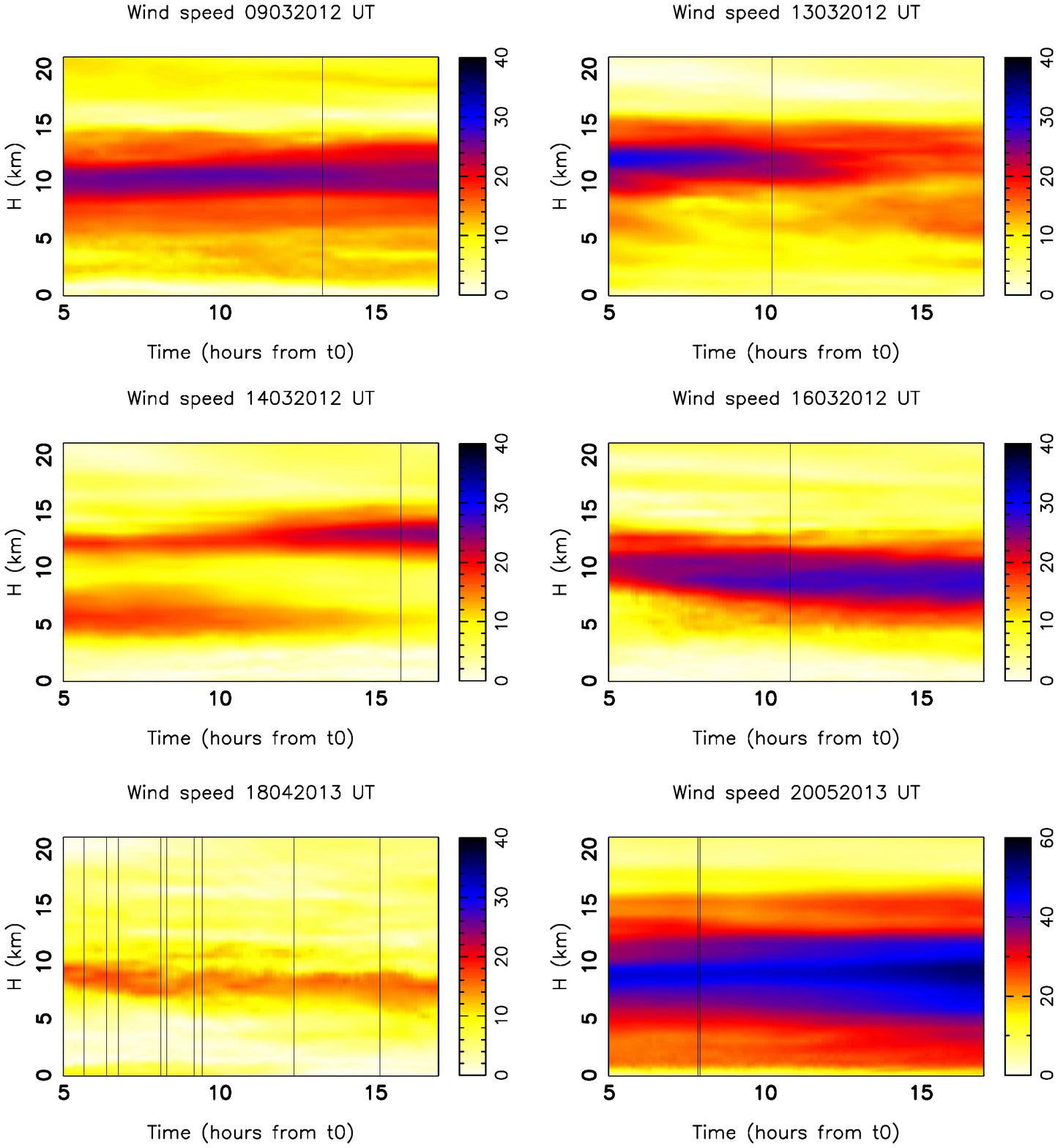} &
    \includegraphics[width = 0.48\linewidth]{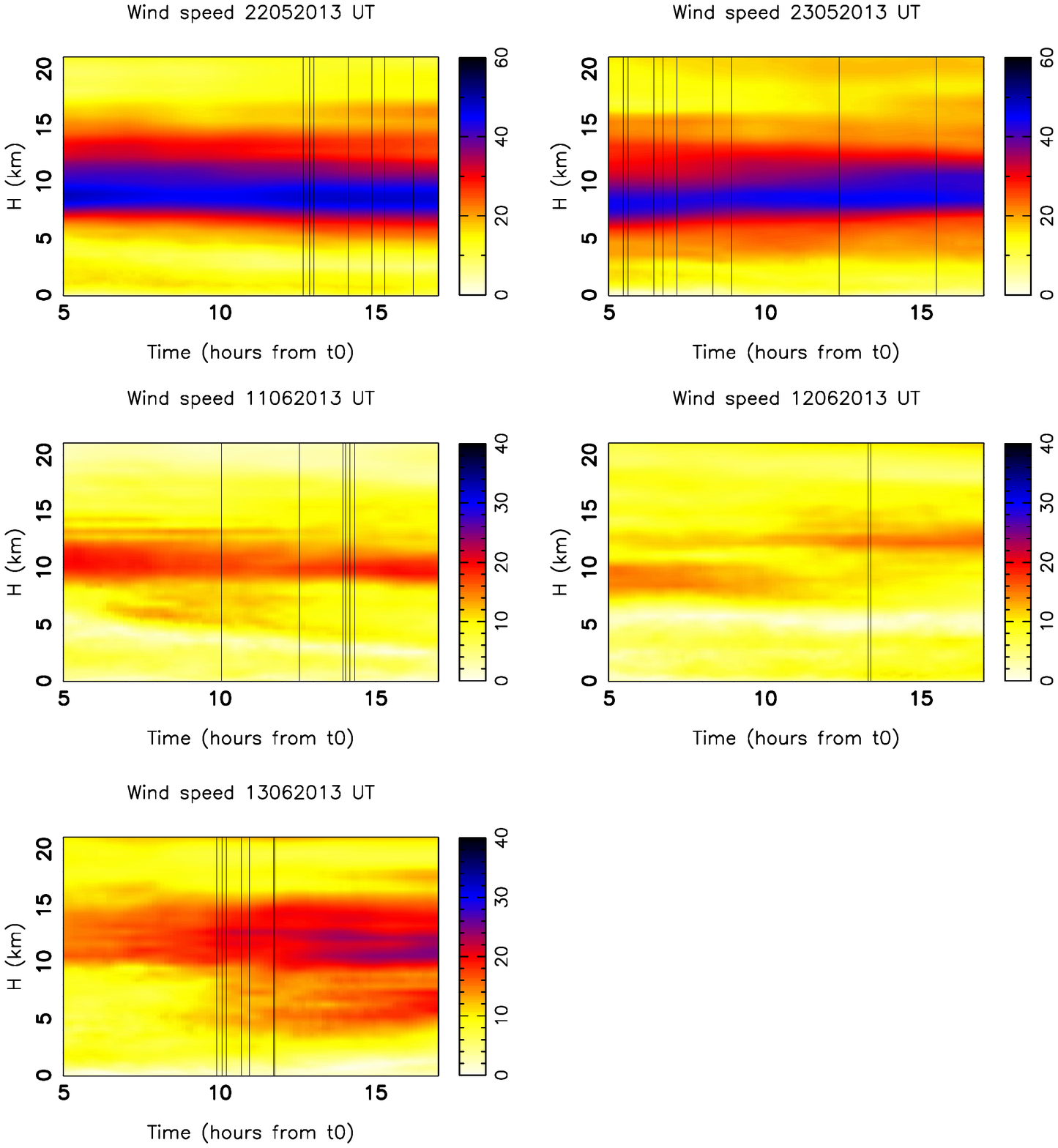}
    \end{tabular}
      \caption[] {Temporal evolution of the wind speed vertical profiles predicted by the model between 18:00 UT of the day before up to 11:00 UT of the selected day. This corresponds to the [14:00 - 07:00] range in local time (LT) (when there is no legal hour). Figures display only the nightly period i.e. the first five hours are discarded. Black vertical lines are the instants for which GeMS measurements are also available. On the y-axis the altitude above the ground (km), on the x-axis the time from the beginning of the simulations (hours). Table of color represent the wind speed intensity in m$\cdot$s$^{-1}$ units.}
  \label{fig:windprofile1}
\end{figure}

We simulated 11 nights, each one starting at at 18:00~UT of the day before and forced each 6 hours with analyses from the European Centre for Medium Range Weather Forecasts (ECMWF). Simulations finish at 11:00~UT of the simulated day (for a total duration of 17 hours). The vertical profiles extracted from the model computations are available each 2 minutes. 
Fig. \ref{fig:windprofile1} shows the temporal evolution of the wind speed vertical profiles during the 11 nights for which GeMS telemetry data were available. 

\begin{figure}
     \begin{tabular}{cc} 
  \includegraphics[width = 0.48\linewidth]{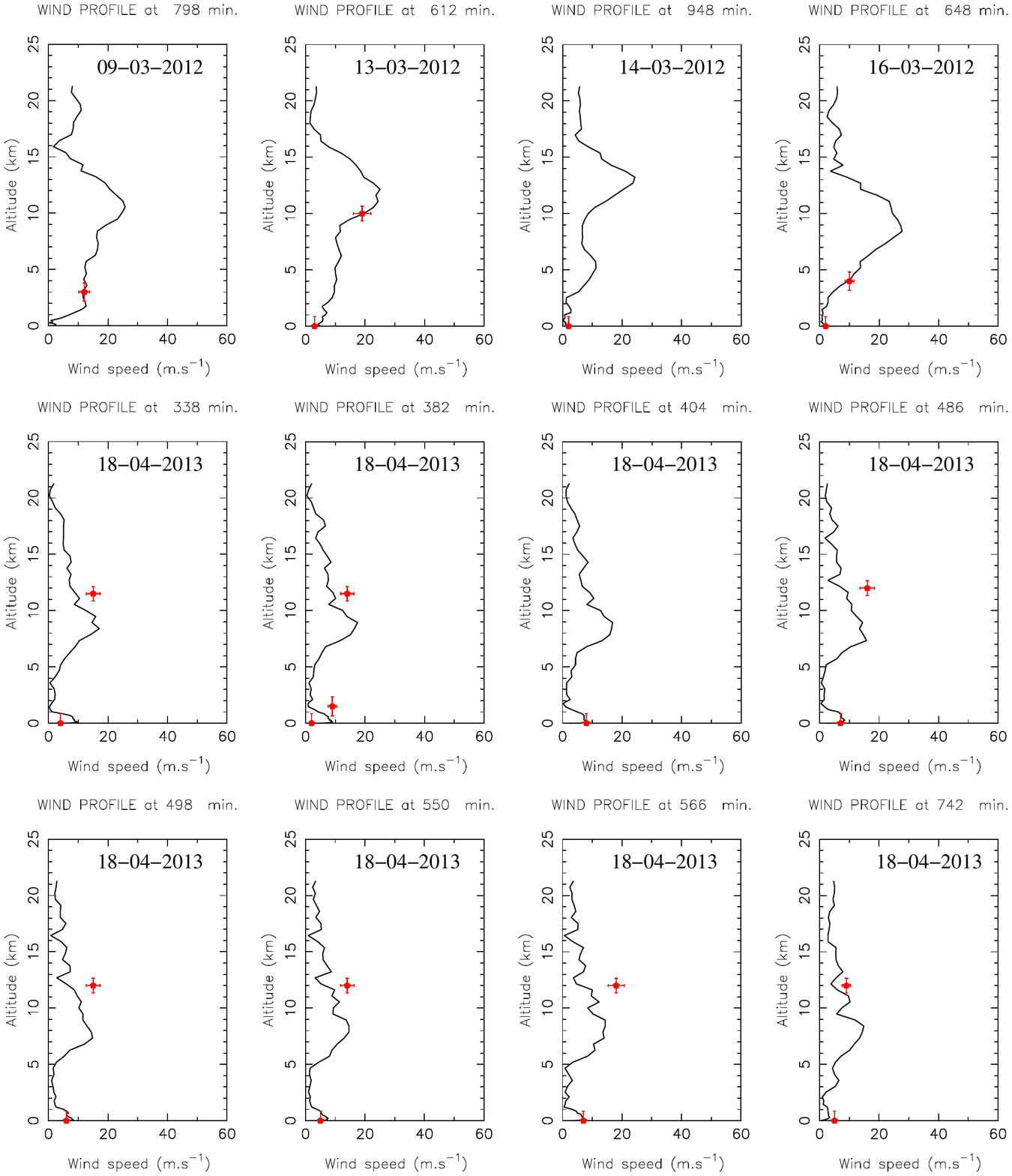} &
    \includegraphics[width = 0.48\linewidth]{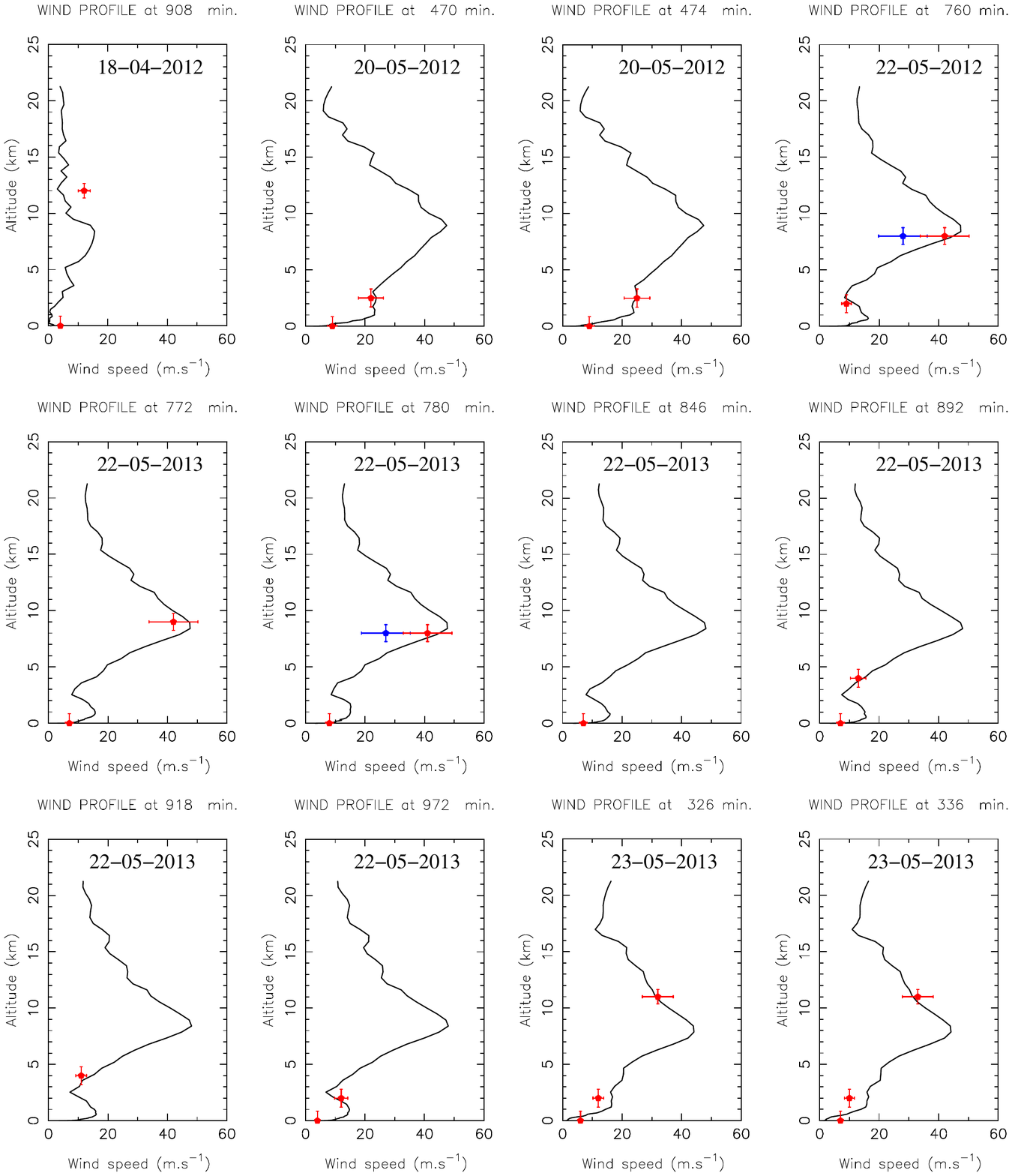}
    \end{tabular}
     \caption[] {Wind speed vertical profiles as reconstructed by the Meso-NH model (black lines) and measured by GeMS (red points). Blue points are estimations obtained without knowing the Meso-Nh predictions. Nights covered are 09/03/2012, 13/03/2012, 14/03/2012, 16/03/2012, 18/03/2012, 20/05/2013, 22/05/2013, 23/05/2013}
  \label{fig:res1}
\end{figure}

\begin{figure}
    \begin{tabular}{cc} 
  \includegraphics[width = 0.48\linewidth]{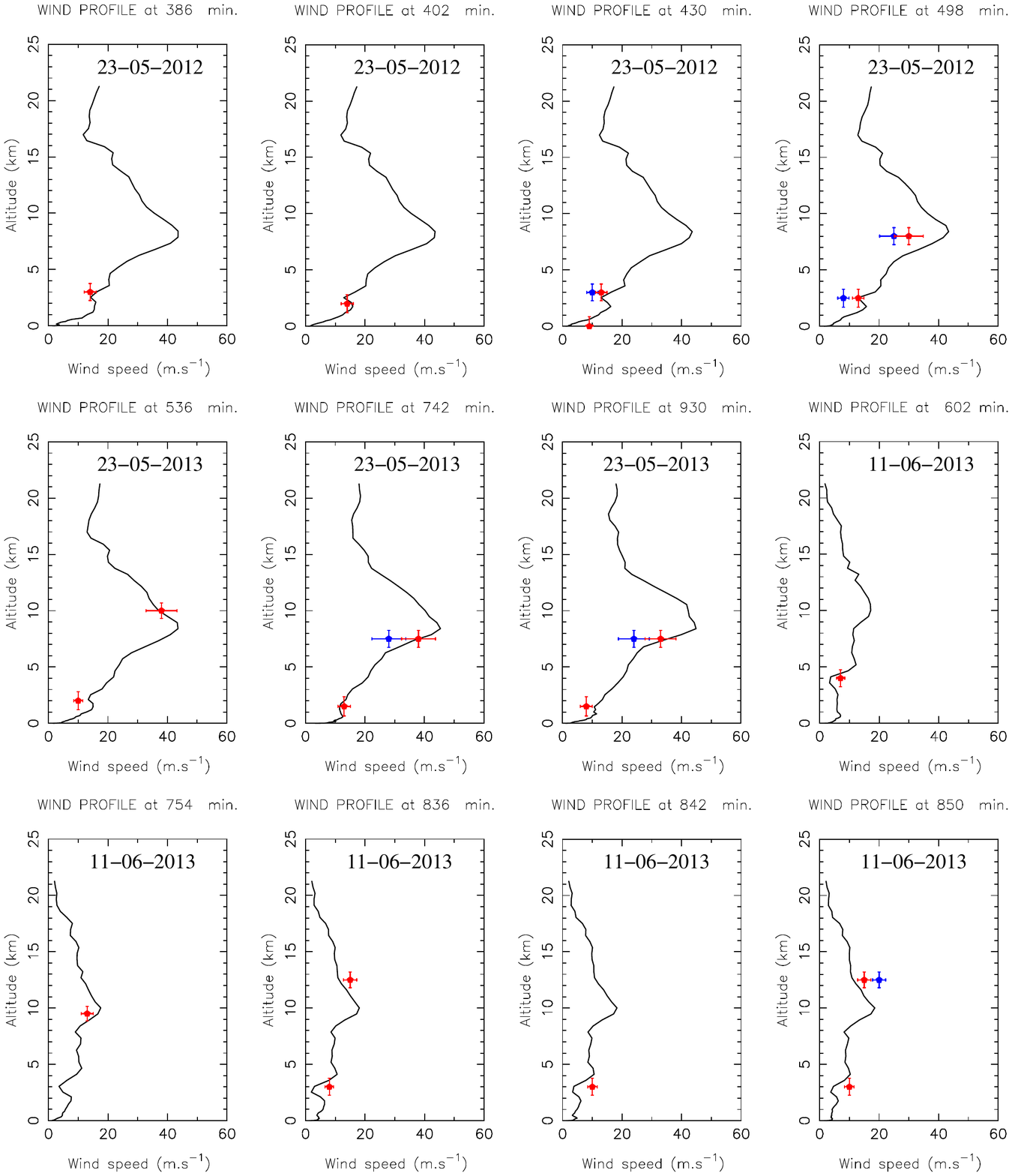} &
    \includegraphics[width = 0.48\linewidth]{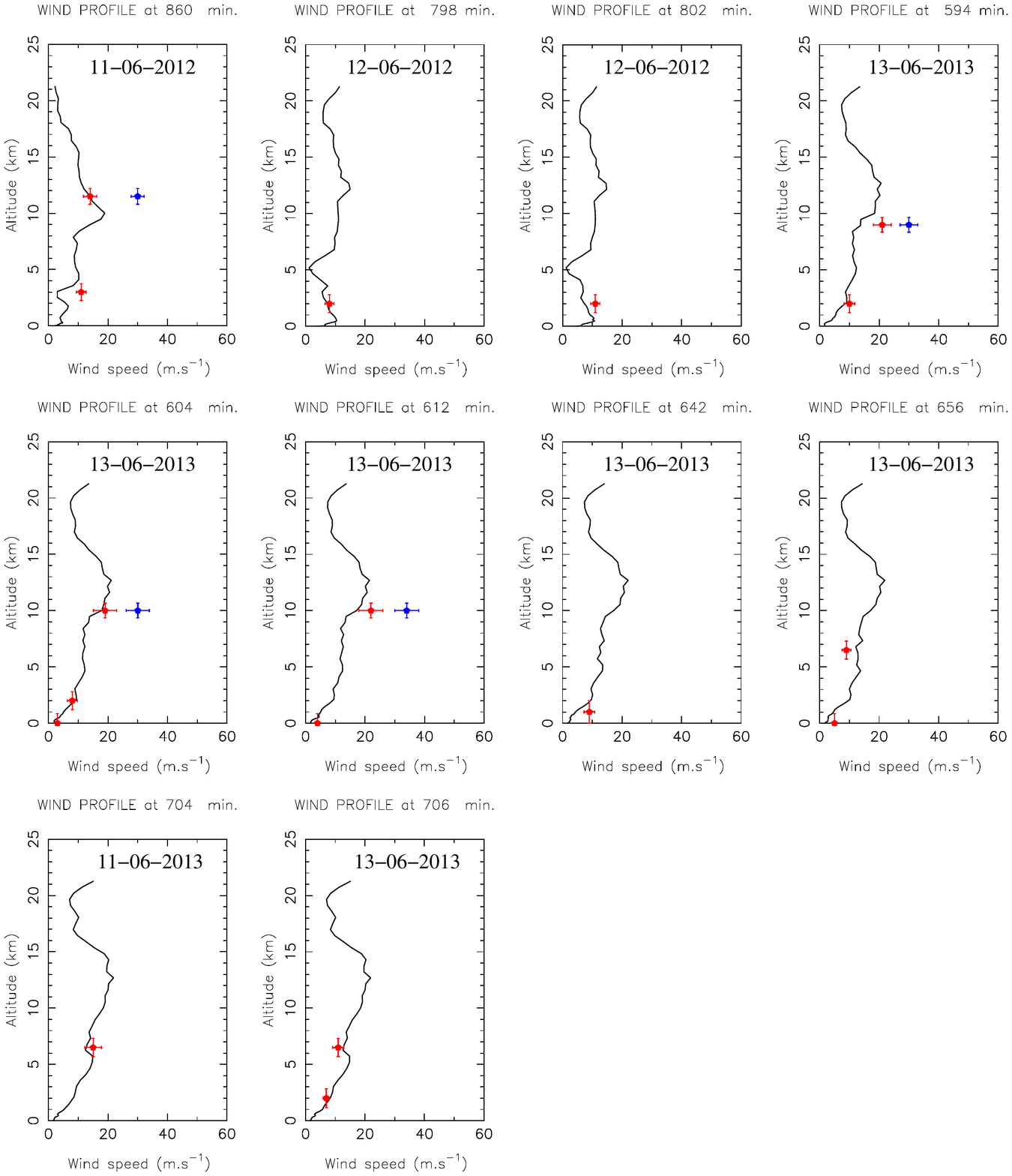}
    \end{tabular}
      \caption[] {As Fig.\ref{fig:res1} - continuing with nights 23/05/2013, 11/06/2013, 12/06/2013, 13/06/2013}
  \label{fig:res2}
\end{figure}


\section{Results} 
\label{sec:results}  

In Fig.~\ref{fig:res1} to \ref{fig:res2} are shown the wind speed vertical profiles as reconstructed by the model Meso-NH (black lines) as well as the GeMS wind speed measurements (red points). The errors bars are calculated considering the uncertainty on the vertical resolution (y-axis) and the dispersion of successive measurements (x-axis). We highlight that the temporal frequency of the model outputs  is 2 min. This means that the temporal coincidence between simulations and measurements is within 2 min. Looking at these results we can retrieve that, overall, there is a very good agreement between the model and GeMS wind speed measurement. For instance, the night of the 22th of May 2013 (Fig. \ref{fig:res1} - right panels) a strong wind shear is predicted by the model around 10km in altitude, which is also perfectly detected by the GeMS data. This is also the case for the night of the 23th of May 2013 (Fig. \ref{fig:res2} - left panels), where a very good match between the prediction and the actual measurements is seen. Another interesting example, good illustration of the correlation between the model and the actual measurements, is for the night of the 13th of June 2013 (Fig. \ref{fig:res2}). \\
We note however that, such good estimates are obtained when both GeMS and Meso-Nh are available simultaneously. Without Meso-Nh, GeMS prediction can, in a few cases, provide not negligible relative errors (see blue points). The reason of the better GeMS data (red points) when Meso-Nh predictions are available is due to the fact that, for some cases, looking at the predictions can actually help to identify a layer moving at a certain velocity at a certain height above the ground. Indeed, there are some layers that are very difficult to see in the GeMS data, but guessing where to look at, based on the predictions, can help a lot to detect them. This is an important point, because if we assume that the predictions are right, then one could use the predictions to look for the correlations peaks in the GeMS wind profiler method. Using the predictions may help to detect more peaks, and then combining both methods might be an interesting way to get the useful information for the tomography.\\

To summarize, if GeMS could be supported by an automated system reading the wind speed from both GeMS and Meso-Nh, one could obtain the double advantage to have (i) a much better 
quantitative estimate of the wind speed from GeMS itself, (ii) to have a total temporal coverage of the whole night with a high temporal frequency and (iii) to have a wind speed forecast, the information from Meso-Nh being available in advance (for example at the beginning of the night).

\section{Conclusions} 
\label{sec:perspectives}
  
We have presented a first comparison between a model predicted wind speed estimation and actual measurements obtained with GeMS. The model can reconstruct the full wind speed distribution over the telescope location. These predictions are then compared with wind speed profiles measured with GeMS. We found that GeMS show frequently a very good level of accuracy. Besides we also note that in a few cases, the presence of simultaneous information from Meso-Nh can help in limiting errors in the procedure used by GeMS to retrieve wind speed estimates or simply in detecting 
wind estimates in layer showing a weak turbulence level.  We conclude, therefore, that the optimized solution might be an automatic systems that is able to read simultaneously both estimates. This comparison between model predictions and GeMS estimates is a first step of a more global effort toward a reliable C$_{N}^{2}$ and wind speed estimation / prediction. Future work will include:
\begin{enumerate}
\item Using a significantly larger data set, and perform a statistical analysis of the matching between model and data.
\item A comparison of the C$_{N}^{2}$ profiles measured by GeMS (the Gemini MCAO systems) with those provided by an independent vertical profiler located nearby:  a MASS (Multi-Aperture Scintillation and Ranging) and a DIMM (Differential Image Motion Monitor) running simultaneously. 
\item A comparison of the C$_{N}^{2}$ profiles measured by GeMS with those predicted by the Meso-NH model.
\item A comparison of the wind direction measured by GeMS with those predicted by the Meso-NH model.
\end{enumerate}

Deploying a Meso-NH prediction tool at the Gemini observatory is also considered.

\acknowledgments     
 Benoit Neichel acknowledges the French ANR program WASABI - ANR-13-PDOC- 0006-01. \\
Based on observations obtained at the Gemini Observatory, which is operated by the Association of Universities for Research in Astronomy, Inc., under a cooperative agreement with the NSF on behalf of the Gemini partnership: the National Science Foundation (USA), the National Research Council (Canada), CONICYT (Chile), the Australian Research Council (Australia), Ministerio da Ciencia, Tecnologia e Inovaccao (Brazil) and Ministerio de Ciencia, Tecnologia e Innovacion Productiva (Argentina). Part of the numerical simulations have been run on the HPCF cluster of the European Center for Medium Range Weather Forecasts (ECMWF) using sources of the Project SPITFOT.


\bibliography{ao,gems}   

\begin{thebibliography}{10}

\bibitem{cortes2012atmospheric}
Cort{\'e}s, A., Neichel, B., Guesalaga, A., Osborn, J., Rigaut, F., and Guzman,
  D., ``Atmospheric turbulence profiling using multiple laser star wavefront
  sensors,'' {\em Monthly Notices of the Royal Astronomical Society}~{\bf
  427}(3),  2089--2099 (2012).

\bibitem{elena2013meso}
Masciadri, E., Lascaux, F., and Fini, L., ``Mose: operational forecast of the
  optical turbulence and atmospheric parameters at european southern
  observatory ground-based sites - i. overview and vertical stratification of
  atmospheric parameters at 0-20 km,'' {\em Monthly Notices of the Royal
  Astronomical Society}~{\bf 436}(3),  1968--1985 (2013).

\bibitem{roddier1981effects}
Roddier, F., ``The effects of atmospheric turbulence in optical astronomy,''
  {\em Progress in optics.}~{\bf 19},  281--376 (1981).

\bibitem{foy1985feasibility}
Foy, R. and Labeyrie, A., ``Feasibility of adaptive telescope with laser
  probe,'' {\em Astron. Astrophys}~{\bf 152},  L29--L31 (1985).

\bibitem{wizinowich2012progress}
Wizinowich, P., ``Progress in laser guide star adaptive optics and lessons
  learned,'' in [{\em Proceedings of SPIE}{\nolinebreak\hspace{0.1em}]},   {\bf
  8447}, International Society for Optics and Photonics (2012).

\bibitem{tallon1990adaptive}
Tallon, M. and Foy, R., ``Adaptive telescope with laser probe-isoplanatism and
  cone effect,'' {\em Astronomy and Astrophysics}~{\bf 235},  549--557 (1990).

\bibitem{johnston1994analysis}
Johnston, D.~C. and Welsh, B.~M., ``Analysis of multiconjugate adaptive
  optics,'' {\em JOSA A}~{\bf 11}(1),  394--408 (1994).

\bibitem{ellerbroek1994first}
Ellerbroek, B.~L., ``First-order performance evaluation of adaptive-optics
  systems for atmospheric-turbulence compensation in extended-field-of-view
  astronomical telescopes,'' {\em JOSA A}~{\bf 11}(2),  783--805 (1994).

\bibitem{fusco2011mcao}
Fusco, T., Conan, J.-M., Rousset, G., Mugnier, L., and Michau, V., ``Optimal
  wave-front reconstruction strategies for multiconjugate adaptive optics,''
  {\em JOSA A}~{\bf 18}(10),  2527--2538 (2001).

\bibitem{neichel2008josa}
Neichel, B., Fusco, T., and Conan, J.-M., ``Tomographic reconstruction for
  wide-field adaptive optics systems: Fourier domain analysis and fundamental
  limitations,'' {\em JOSA A}~{\bf 26}(1),  219 (2008).

\bibitem{conan2001mcao}
Conan, J.-M., Le~Roux, B., Bello, D., Fusco, T., and Rousset, G., ``Optimal
  reconstruction in multiconjugate adaptive optics,'' in [{\em Beyond
  conventional adaptive optics}{\nolinebreak\hspace{0.1em}]},   {\bf 58},  209
  (2001).

\bibitem{toko2001tomo}
Tokovinin, A. and Viard, E., ``Limiting precision of tomographic phase
  estimation,'' {\em JOSA-A}~{\bf 18}(8),  873--882 (2001).

\bibitem{miska2001tomo}
LeLouarn, M. and Tallon, M., ``Analysis of modes and behavior of a
  multiconjugate adaptive optics system,'' {\em JOSA-A}~{\bf 19}(8),  912--925
  (2002).

\bibitem{fusco99phase}
Fusco, T., Conan, J., Michau, V., Mugnier, L., and Rousset, G., ``Efficient
  phase estimation for large-field-of-view adaptive optics,'' {\em Opt.
  Letter}~{\bf 24}(1),  1472--1474 (1999).

\bibitem{costille2012cn2}
Costille, A. and Fusco, T., ``Impact of cn2 profile on tomographic
  reconstruction performance: application to e-elt wide field ao systems,'' in
  [{\em Proceedings of SPIE}{\nolinebreak\hspace{0.1em}]},   {\bf 8447},
  844757 (2012).

\bibitem{gavel2002prediction}
Gavel, D.~T. and Wiberg, D., ``Towards strehl-optimizing adaptive optics
  controllers,'' in [{\em Proceedings of SPIE}{\nolinebreak\hspace{0.1em}]},
  {\bf 4839},  890--901 (2002).

\bibitem{leroux2004prediction}
Le~Roux, B., Conan, J.-M., Kulcsar, C., Raynaud, H.-F., Mugnier, L.~M., and
  Fusco, T., ``Optimal control law for classical and multiconjugate adaptive
  optics,'' {\em JOSA A}~{\bf 21}(A),  1261--1276 (2004).

\bibitem{hinnen2010prediction}
Hinnen, K., Verhagen, M., and Doelman, N., ``Exploiting the spatiotemporal
  correlation in adaptive optics using datadriven h2-optimal control,'' {\em
  JOSA A}~{\bf 24}(A),  1714--1725 (2010).

\bibitem{poyneer2007prediction}
Poyneer, L., Macintosh, B.~A., and Véran, J.-P., ``Fourier transform
  wavefront control with adaptive prediction of the atmosphere,'' {\em JOSA
  A}~{\bf 24}(A),  2645--2660 (2007).

\bibitem{ammons2012tomo}
Ammons, S.~M., Poyneer, L., Gavel, D.~T., Kupke, R., Max, C.~E., and Johnson,
  L., ``Evidence that wind prediction with multiple guide stars reduces
  tomographic errors and expands moao field of regard,'' in [{\em Proceedings
  of SPIE}{\nolinebreak\hspace{0.1em}]},   {\bf 8447},  84471U (2012).

\bibitem{rigaut2014gems}
Rigaut, F., Neichel, B., Boccas, M., and et~al., ``Gemini multiconjugate
  adaptive optics system review - i. design, trade-offs and integration,'' {\em
  MNRAS}~{\bf 437}(3),  2361--2375 (2014).

\bibitem{neichel2014gems}
Neichel, B., Rigaut, F., Vidal, F., and et~al., ``Gemini multiconjugate
  adaptive optics system review - ii. commissioning, operation and overall
  performance,'' {\em MNRAS}~{\bf 440}(2),  1002--1019 (2014).

\bibitem{mcgregor2004gemini}
McGregor, P., Hart, J., Stevanovic, D., Bloxham, G., Jones, D., Van~Harmelen,
  J., Griesbach, J., Dawson, M., et~al., ``Gemini south adaptive optics imager
  (gsaoi),'' in [{\em Proceeding of SPIE}{\nolinebreak\hspace{0.1em}]},   {\bf
  5492},  1033--1044, International Society for Optics and Photonics (2004).

\bibitem{elston2003performance}
Elston, R., Raines, S.~N., Hanna, K.~T., Hon, D.~B., Julian, J., Horrobin, M.,
  Harmer, C.~F., and Epps, H.~W., ``Performance of the flamingos near-ir
  multi-object spectrometer and imager and plans for flamingos-2: a fully
  cryogenic near-ir mos for gemini south,'' in [{\em Proceeding of
  SPIE}{\nolinebreak\hspace{0.1em}]},   {\bf 4841},  1611--1624 (2003).

\bibitem{wilson2002slodar}
Wilson, R., ``Slodar: measuring optical turbulence altitude with a
  shack-hartmann wavefront sensor,'' {\em Monthly Notices of the Royal
  Astronomical Society}~{\bf 337}(1),  103--108 (2012).

\bibitem{gilles2010cn2}
Gilles, L. and Ellerbroek, B.~L., ``Real-time turbulence profiling with a pair
  of laser guide star shack-hartmann wavefront sensors for wide-field adaptive
  optics systems on large to extremely large telescopes,'' {\em JOSA A}~{\bf
  27}(11),  A76 (2010).

\bibitem{osborn2012neuro}
Osborn, J., de~Cos~Juez, F.~J., Guzman, D., and et~al., ``Using artificial
  neural networks for open-loop tomography,'' {\em Optics Express}~{\bf 20}(3),
   2420 (2012).

\bibitem{wang2008cn2}
Wang, L., Schock, M., and Chanan, G., ``Atmospheric turbulence profiling with
  slodar using multiple adaptive optics wavefront sensors,'' {\em Applied
  Optics}~{\bf 47}(11),  1880--1892 (2008).

\bibitem{andres2014cn2}
Guesalaga, A., Neichel, B., Cortes, A., Bechet, C., and Guzman, D., ``Using the
  cn2 and wind profiler method with wide-field laser-guide-stars adaptive
  optics to quantify the frozen-flow decay,'' {\em Monthly Notices of the Royal
  Astronomical Society}~{\bf 440}(3),  1925--1933 (2014).

\bibitem{valenzuela2014spie}
Valenzuela, J., Guesalaga, A., C., B., and et~al., ``Comparison of two
  turbulence profilers applied to eso's adaptive optics facility,'' in [{\em
  Proceeding of SPIE}{\nolinebreak\hspace{0.1em}]},   {\bf 9148-66},
  International Society for Optics and Photonics (2014).

\bibitem{ignacio2014spie}
Rodriguez, I., Neichel, B., Guesalaga, A., and et~al., ``Statistics of
  atmospheric turbulence at cerro pachon using the gems profiler,'' in [{\em
  Proceeding of SPIE}{\nolinebreak\hspace{0.1em}]},   {\bf 9148-240},
  International Society for Optics and Photonics (2014).

\bibitem{lafore1998}
Lafore, J. and al., ``The meso-nh atmospheric simulation system. part i:
  adiabatic formulation and control simulations. scientific objectives and
  experimental design,'' {\em Ann. Geophys.}~{\bf 60},  90--109 (1998).

\bibitem{masciadri1999}
Masciadri, E., Vernin, J., and Bougeault, P., ``3d mapping of optical
  turbulence using an atmospheric numerical model. i. a useful tool for the
  ground-based astronomy,'' {\em Astronomy and Astrophysics Supplementary
  Series}~{\bf 137},  185--202 (1999).

\bibitem{masciadri2004}
Masciadri, E., Avila, R., and Sanchez, L., ``Statistic reliability of the
  meso-nh atmospherical model for 3d c$_{N}^{2}$ simulations,'' {\em Revista
  Mexicana de Astronomia y Astrofisica}~{\bf 40},  3--14 (2004).

\bibitem{masciadri2006}
Masciadri, E. and Egner, S., ``First seasonal study of optical turbulence with
  an atmospheric model,'' {\em PASP}~{\bf 118},  1604--1619 (2006).

\bibitem{hagelin2010}
Hagelin, S., Masciadri, E., and Lascaux, F., ``Wind speed vertical distribution
  at mt. graham,'' {\em MNRAS}~{\bf 407},  2230--2240 (2010).

\bibitem{hagelin2011}
Hagelin, S., Masciadri, E., and Lascaux, F., ``Characterization of the optical
  turbulence at mt. graham using the meso-nh model,'' {\em MNRAS}~{\bf 412},
  2695--2706 (2011).

\bibitem{lascaux2012}
Lascaux, F., Masciadri, E., and Hagelin, S., ``Mesoscale optical turbulence
  simulations above dome c, dome a and south pole,'' {\em MNRAS}~{\bf 411},
  693--704 (2011).

\end{thebibliography}
\bibliographystyle{spiebib}   

\end{document}